%% file: mainIEEE.tex
\documentclass[conference]{IEEEtran}
\IEEEoverridecommandlockouts
\usepackage{cite}
\usepackage{booktabs}
\usepackage{amsmath,amssymb,amsfonts}
\usepackage{algorithmic}
\usepackage{hyperref}
\usepackage{graphicx}
\graphicspath{ {./pictures/} }
\usepackage{textcomp}
\usepackage{xcolor}
\usepackage{multirow}
\definecolor{babyblueeyes}{rgb}{0.63, 0.79, 0.95}
\usepackage[most]{tcolorbox} 
\newtcolorbox{barquote}{%
    colback=white,
    grow to right by=-1mm,
    grow to left by=-1mm, 
    boxrule=0pt,
    boxsep=0pt,
    breakable,
    enhanced jigsaw,
    borderline west={4pt}{0pt}{babyblueeyes},
}
   
\begin{document}

\title{DevServOps: DevOps For Product-Oriented Product-Service Systems}

%
\author{\IEEEauthorblockN{Anas Dakkak}
\IEEEauthorblockA{\textit{Ericsson AB} \\
Stockholm, Sweden \\
anas.dakkak@ericsson.com}
\and
\IEEEauthorblockN{Jan Bosch}
\IEEEauthorblockA{\textit{Chalmers University of Technology} \\
Göteborg, Sweden \\
jan.bosch@chalmers.se}
\and
\IEEEauthorblockN{Helena Holmström Olsson}
\IEEEauthorblockA{\textit{Malmö University} \\
Malmö, Sweden \\
helena.holmstrom.olsson@mau.se}
}

\maketitle
\begin{abstract}
    \input{content/0abstract}

\end{abstract}


\input{content/1intro}

\input{content/2background}

\input{content/3method}

\input{content/4context}
\input{content/5results}

\input{content/6discussion}
\input{content/7conclusion}



\bibliographystyle{IEEEtran.bst}
\bibliography{bib.bib}
\end{document}

%% file: content/0abstract.tex
For companies developing web-based applications, the \textit{Dev} and the \textit{Ops} refer to different groups with either operational or development focus. Therefore, DevOps help these companies streamline software development and operations activities by emphasizing the collaboration between the two groups. However, for companies producing software-intensive products, the \textit{Ops} would refer to customers who use and operate the product. In addition, companies producing software-intensive products do not only offer products to customers but rather Product Service Systems (PSS), where product-related services play a key role in ensuring customer satisfaction besides their significant revenue contribution. Thus, the context of product-oriented PSS is very different from web-based applications, making it difficult to apply DevOps without considering the role of the services. Therefore, based on a two years participant observation case study conducted at a multinational telecommunications systems provider, we propose a new and novel approach called Development-Services-Operations (DevServOps) which incorporates services as a key player facilitating an end-to-end software flow toward customers in one direction and feedback toward developers in the other direction. Services become the glue that connects the Dev and the Ops, achieved by providing internal services to increase the precision of the development organization and external services to increase the speed of deployment and new content adoption on the customers' side.

%% file: content/1intro.tex
\section{Introduction}

Since its advent, DevOps has been gaining considerable popularity in the software industry, especially among companies developing web-based applications \cite{lwakatare2019devops}. At its core, DevOps emphasizes collaboration between the teams developing the software on the one hand and the ones operating the software and its infrastructure, on the other hand, to streamline the activities from code development until its realization in production. \cite{luz2019adopting}. As a consequence, the boundary between development and operations becomes blurry as the notion of ``handover to operations'' is replaced by ``you build it, you run it'' as used in Netflix \cite{ebert2022devops}, or by enforcing a culture of personal responsibility as used in Facebook \cite{feitelson2013development}. While DevOps can be realized in multiple ways depending if the web application is deployed on-prem, on hybrid cloud, or on cloud-based infrastructure \cite{macarthy2020empirical}, DevOps in the context of web-based applications relies on two main enablers: first, there are no other interfaces between the Dev and the Ops; thus, a streamlined software flow can be achieved by tearing down the wall between development and operations. Second, the customers are not part of the collaboration, as software deployment, operations, and even online experimentation, are largely transparent activities to them. 

However, these two enablers are often missing in the case of software-intensive products, such as cars, airplanes, and telecommunications network equipment, as the context is fundamentally different from web-based applications. First, companies producing software-intensive products do not provide products only to their customers but Product Service Systems (PSS), a bundle of products and services aimed at fulfilling customers' needs \cite{gebauer2016services, baines2009servitization}. A widely adopted approach of PSS is when companies sell products to their customers; thus, the ownership of the product moves to the customer while the company provides product-related services such as maintenance, upgrades, optimization, and support \cite{baines2007state}. This is referred to as product-oriented PSS, where services aim to guarantee the product's functionality and durability. Therefore, services provided by product vendors play a key role in ensuring customer satisfaction and loyalty, besides their significant contribution to the company's revenue \cite{gebauer2016services}. To deliver product-related services, product-oriented PSS companies often establish dedicated service units interfacing with customers and working closely with them \cite{brosig2022selling, oliva2012separate}. Second, deploying a new software version into customers' products is not transparent activity as products often become unavailable during software upgrades or, in some cases, require pre-deployment lab validation in a customer-like environment \cite{lwakatare2016towards}.

Therefore, in product-oriented PSS context, the \textit{Ops} of DevOps would refer to the customer who operates and uses the product, while the \textit{Dev} would refer to the software development teams within the R\&D organization of the product's vendor. Despite their critical role for customers and product-oriented PSS companies, this leaves services out of the picture. Besides, While PSS and DevOps have been researched extensively, they have been addressed as separate topics. DevOps empirical studies have been focusing on the product side by investigating the challenges of adopting DevOps and the R\&D organizations' evolution models \cite{lwakatare2016towards, olsson2014climbing}, while studies addressing PSS have focused on requirement engineering and design guidelines, among other topics \cite{song2017requirement, muto2015guideline}.

Therefore, this paper investigates the role of services with DevOps in a product-oriented PSS context. In addition, it conceptualizes the Development-Services-Operations (DevServOps) as a novel approach that incorporates services with DevOps to streamline the end-to-end software flow in a product-oriented PSS context. DevServOps emphasizes the role of services as the glue connecting the Dev represented by the vendor's product development and the Ops represented by customers. The DevServOps has been derived from a two-year participant observation case study that combines qualitative and qualitative data at a multinational telecommunication company that provides products and services to hundreds of mobile network operators globally.  

The contribution of this paper is three-fold. First, it explores the role of services with DevOps in the context of PSS, thus, addressing an existing software engineering research gap. Second, it coins the term \textit{DevServOps}, which emphasizes the importance of services with DevOps in the context of PSS. By bringing the service perspective to DevOps, we provide a new and holistic approach addressing the context of product-oriented PSS companies and ensuring customers' satisfaction. Third, this case study is conducted at one of the largest telecommunications vendors in the world with established frequent releases of complex embedded software and close collaboration with customers; thus, the paper provides practical insights into the role of services with DevOps and how DevServOps is applied in practice.

The remainder of the paper is organized as follows. Section \ref{DevServOps_background} provides an overview and a literature review of DevOps and PSS systems. Section \ref{devservops_research_method} details the research method used in this study, including data collection, analysis, and threats to validity. Section \ref{devservops_context} provides a background about the case study company, while Sections \ref{devservops_results} present the empirical results. Section \ref{devservops_discussion} discusses the results and presents the DevServOp concept. Finally, section \ref{devservops_conclusion} concludes the paper.

%% file: content/2background.tex
\section{Background} \label{DevServOps_background}


In response to the increasing market competition, commoditization of products, and reduced product profit, product-oriented companies started embracing services as a new revenue source and a competitiveness differentiator \cite{tan2010service, baines2009servitization}. As a result, the notion of Product Service Systems has evolved as a marketable set of products and services capable of jointly fulfilling a user’s need \cite{goedkoop1999product}. PSS can be classified into a product-, use- or result-oriented PSS  \cite{salwin2020state, gebauer2016services}. In \textit{product-oriented PSS}, the customer purchases and owns the product while the vendor provides additional services, such as maintenance, upgrade, or support. Product-oriented PSS is a widely adopted approach in many companies.  In \textit{use-oriented PSS}, the product is owned and serviced by the vendor who sells its function to the customer, for example, by leasing. Third, \textit{result oriented PSS} where the vendor owns and services the product but sells the product's result, per usage, to customers.

\subsection{Product-oriented Product service systems}

Product-oriented PSS companies provide products and product-related services to customers. Products represent the tangible goods capable of ``falling on your toes" and fulfilling a user’s needs \cite{baines2007state}. While the term refers to the physical aspect of the product, Berkovich et al. \cite{berkovich2011requirements} highlighted that PSS represents an \textit{integrated bundle of hardware, software, and service elements to solve customers’ problems}. Other researchers conceptualize new terms such as \textit{Industrial Software-Product-Service Systems} and \textit{smart PSS} to emphasize the role of software, AI/ML, and IoT connectivity \cite{chowdhury2018smart}. Despite the definition and the name, many products are becoming software-intensive as the software is taking over the hardware and electronics as the main component shaping the product's functionality and the driver of innovation \cite{andersson2021software, bosch2021digital}. As a result, new software versions can bring added functionalities and corrections to the product without the need to change the hardware, which opens the possibility to make customers' products continuously evolve throughout their longer hardware lifespan. As a result, companies producing software-intense adopt software development practices stepwise, first by introducing development practices such as scrum and continuous integration, leading to reduced release frequency from years or months to a few weeks \cite{olsson2012climbing}. These development practices are performed internally in the development organization to streamline coding, testing, and integration. However, as companies move to continuous deployment, collaboration with customers becomes critical to extending the software flow to customers' products.  

Services represent the intangible goods, which are activities done on a commercial basis and for an economic value \cite{baines2007state, goedkoop1999product}. Services play a key role on several fronts. On the business front, services contribute significantly to companies' revenues. Szwejczewski et al. \cite{szwejczewski2015product} found that services sales account for 15-52\% of the revenue when investigating the role of services in six product-oriented companies. Similarly, Gebauer et al. \cite{gebauer2016services}, highlighted the significant contribution of services to companies such as IBM, Ericsson, and ABB. On the customer satisfaction front, services are key to increasing customer satisfaction, by, for example reducing faults detection and correction lead time to increase the product's availability \cite{szwejczewski2015product, windley2002delivering}.

Mathieu et al. \cite{mathieu2001product}, classified services into two categories: services supporting the product, such as maintenance, upgrade, and optimization, and services supporting the customer, such as education and learning. Therefore, to deliver these services, product-oriented companies establish a dedicated services organization alongside the product development organization, which is responsible for the services' portfolio development and delivery \cite{brosig2022selling, oliva2012separate}. The service organization has dedicated service staff and teams who work closely with customers to deliver these services. Thus, customer service is often the interface customers interact with once they start using the product. 

\subsection{DevOps and product-oriented PSS}

DevOps was coined by Patrick Debois in 2008 as he argued that a successful introduction of agile requires a tight collaboration between the development and operations teams \cite{debois2008agile}. While there is no widely agreed definition for DevOps \cite{teixeira2020systematic}, several researchers attempted to conceptualize DevOps focusing on web-based applications. For example, Leite et al. \cite{leite2019survey} created a conceptual framework of fundamental concepts of DevOps, which includes four categories: people, process, delivery, and runtime, while Lwakatare et al. \cite{lwakatare2015dimensions} highlighted that automation, measurement, and monitoring are important dimensions of DevOps.

However, PSS has a very different context to web-based applications. Lwakatare et al. \cite{lwakatare2016towards} identified a number of barriers to DevOps adoption in software-embedded systems companies, such as a lack of lab automation and the difficulty in getting information from customers. While this research is based on multiple case studies performed in industrial companies, it is concerned with the products only and didn't address the role of services with DevOps. Similarly, Srinivasan et al. \cite{srinivasan2019structure}, conducted a case study at Swisscom AG, which produces a PSS solution composed of telecommunications equipment deployed in customers' homes, allowing them to consume different services, such as TV, video on demand, etc. The authors identified the enablers for DevOps, which they classified into process, organization, and product enablers. While the study provides insights into how DevOps is implemented in PSS context, it focuses on software development and does not address the role of service.

%% file: content/3method.tex
\section{Research Method} \label{devservops_research_method}

To explore the role of services with DevOps in the context of product-oriented PSS, we conducted a case study at a multinational telecommunications systems vendor. We chose a case study because it is a suitable research method for the evaluation of software engineering practices in an industrial setting, enabling the researchers to gain an in-depth understanding of the study phenomenon \cite{yin2017case}.

\subsection{Case study company}

The case study company is a multinational telecommunications systems vendor, providing products and services to a wide range of customers, such as mobile service providers, national telecommunications agencies, and enterprises. The case study company has over a hundred thousand employees, of which thirty thousand work in product development while the rest are in services or other supporting functions like sales and administration. Services revenue constitutes 35\% of the total annual revenue, while the remaining is product revenue.

Among many telecommunication systems the case study company produces, we focused on the 5G Radio Access Networks (5G RAN). 5G RAN system is made of a number of interconnected Radio Base Station (RBS), called next-generation NodeB (gNodeB). Each gNodeB provides 5G radio coverage, ranging from several meters to several kilometers. The gNodeB comprises several purpose-built hardware components, such as antennas, baseband processing units, and embedded software. The embedded software plays a key role in the gNodeB as it hosts the majority of the functionalities provided by the gNodeB. 

The case study company and the 5G RAN system are selected for three reasons. First, the 5G RAN system consists of hardware, software, and product-related services. Customers purchase the hardware, subscribe to the software, and contract product-related services independently; thus, the 5G RAN solutions can be considered product-oriented PSS. Second, 5G RAN embedded software is released every second week to a subset of customers who have close collaboration with the case study company, fulfilling two enablers of DevOps; collaboration and frequent releases. Third, the first author is employed by the case study company and works as an R\&D manager and embedded researcher, enabling him to collect and analyze a significant amount of qualitative data from multiple sources while engaging as a participant observer for two years.

\subsection{Data collection and analysis}

This study used qualitative and quantitative data collected over two years, between January 2021 and December 2022. During this time, the service organization in the case study company started an internal project aiming to restructure the services portfolio as a response to the introduction of frequent software releases by the R\&D development organization and an increasing number of customers who would like to establish a closer and more collaborative relationship with the case study company. The first author has been part of the project team and participated in numerous meetings about DevOps and product-related services evolution. These meetings were held weekly with members from the service and R\&D organizations. In addition to meetings participation, the first author has organized and led more than 10 workshops focusing on service evolution with wider participation from other organizational functions such as security, software supply, product introduction, and quality attended. In addition, the authors reviewed more than 20 internal documents describing product-related services and the software development process. These documents were accessible from the internal intranet and were identified by the first author of this study, referred to during meetings or during one of the interviews.

Moreover, quantitative data were also collected and analyzed during the study period. The data originate from three main sources. First, an internal database that contains information about continuous integration, delivery, and release. This database acts as the information bank for all activities related to software development. The second data source is in-service product data collected from customers' networks, including products' software versions and configurations. The third source of information is a services database which includes an inventory of issues triggering a technical investigation. While quantitative data is extensive in size and depth, we extracted, aggregated, and analyzed subsets of these data in light of the objective of this research. To protect the confidentiality of the case study company, absolute numbers presented in this study are normalized and displayed in percentage. 

We followed the guidelines suggested by Runeson et al. \cite{runeson2009guidelines} for software engineering case studies. The steps followed are case study design, preparation for data collection, collecting evidence, analysis of the collected data, and reporting. Since this study was conducted over two years, the data collection and analysis phases were iterative, where we continuously collected and analyzed new data.

\subsection{Validity considerations}

Threats to validity can be classified into four categories: construct, external, and internal validity, in addition to reliability \cite{yin2017case,runeson2009guidelines}. The authors of this paper are all familiar with DevOps practices and product-related services. In addition, the authors had a \textit{prolonged involvement} with the case study company enabling them to gain an in-depth understanding of the case study context. Further, as the first author is employed by the case study company and was a member of the internal project team, there is a risk that his own views and interpretations might influence the data collection and analysis. In addition, We used triangulation with different data sources to minimize the risk of bias and address construct validity and reliability.  

This study is a single case study conducted in a telecommunications company that provides product-oriented PSS. Therefore, we do not claim the generalizability of the results. More empirical studies and research is needed to achieve external validity. However, we believe there are many similarities between the case study company, and other large-scale software-intensive product vendors, especially the ones working in a business-to-business context. 

While this is an explorative case study that does not aim to examine causal relationships, which is the key concern for internal validity, we used \textit{peer debriefing} among the authors to improve the internal validity of the case study research.

%% file: content/4context.tex
\section{The case study company context} \label{devservops_context}

The case study company represents the \textit{Dev} side of DevOps, as it is responsible for developing, testing, and releasing the 5G RAN software, while customers represent the \textit{Ops} side, as they deploy the release and operate the 5G RAN network. Further, the case study company has several customers, each with thousands of gNodeBs in their networks; thus, there are many instances of the \textit{Ops} served by one \textit{Dev} organization. 

\begin{figure}[htbp]
	\begin{center}
	\includegraphics[width=3.2in, height=2.8in]{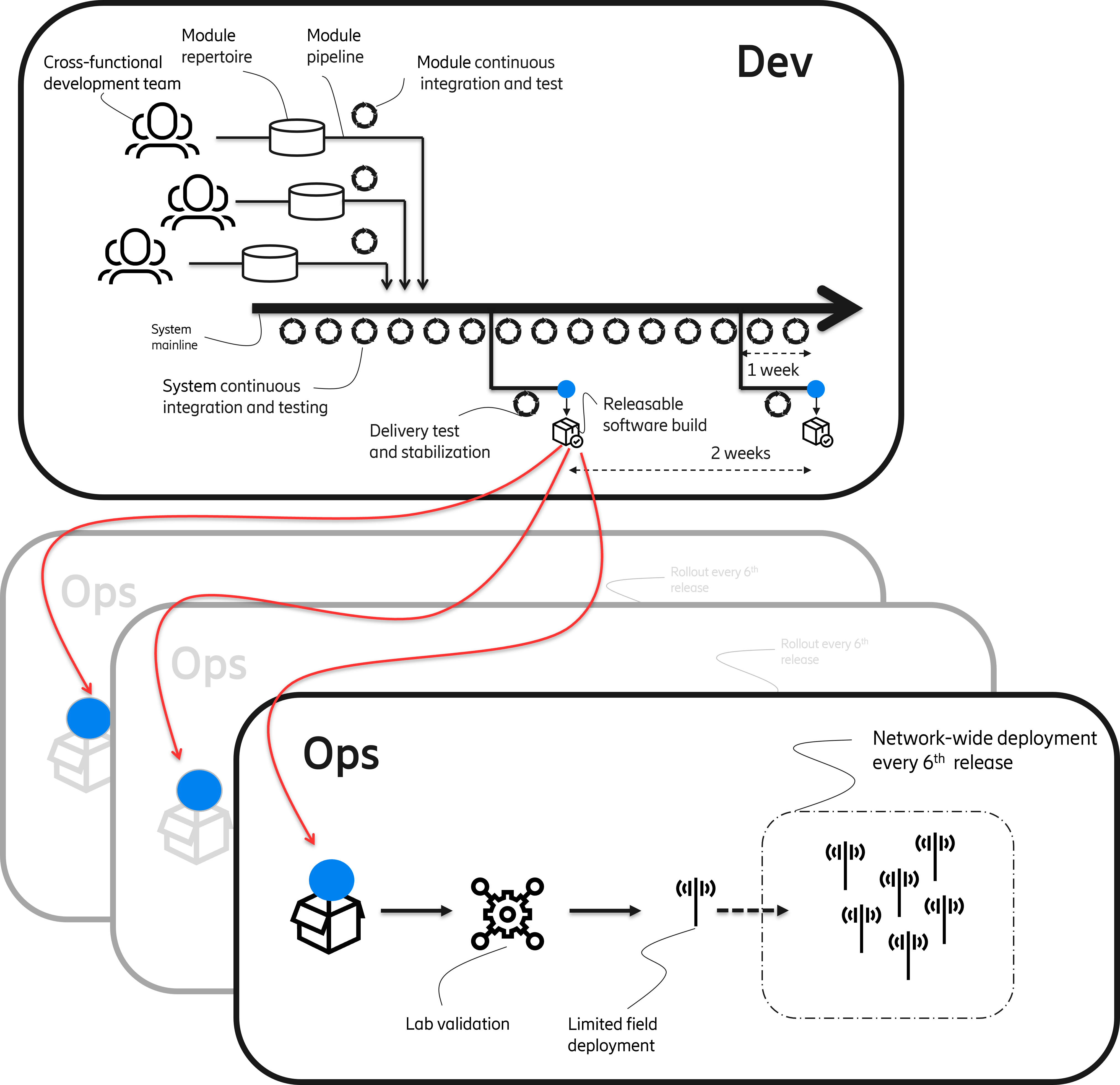}
	\caption{5G RAN software flow}
	\label{fig:DevOpsSWflow}
	\end{center}
\end{figure}

The RAN software is composed of several large components referred to as ``modules''. As illustrated in Figure \ref{fig:DevOpsSWflow}, each module is developed by one or several cross-functional teams. The teams push new code into the modules' repertoire, which triggers a module build and test. Following that, a successfully built and tested module is pushed to the system's mainline, where a system image, composed of all modules, is integrated and continuously tested. As a result of significant investment and improvements in continuous integration over many years, the lead time from a code change until a potentially releasable system image has been shortened from months to a few days. 

Traditionally, the case study company used to release a new major software release every three months, followed by a maintenance period. In 5G RAN, the case study company introduced a calendar-based bi-weekly release to selective customers. These bi-weekly releases are not deployed to the entire network but to a small subset of the product install base, following a similar approach to what we have described in our earlier work \cite{dakkak2022controlled}. These releases are referred to as \textit{CD (Continuous deployment) releases} for two reasons: (1) they are released and deployed in a shorter frequency than major releases (2) these releases contain both new functions and fixes as they are selected from the code's mainline. 

The customers who subscribe to the bi-weekly releases are referred to as \textit{collaborators}, as they represent a small segment of customers who are technology leaders with a strong interest in adopting the latest features to maintain their market-leading positions. Other customers who do not subscribe to the bi-weekly releases and only use major releases are considered \textit{traditional} customers. Thus, traditional customers receive new features and content once every quarter with every new major release, while the collaborators receive new software every second week. After the software is made available to customers, customers download and unpack the software, explore its content and impact by reading the release documents, and test it in a lab before deploying it t the software to a small but representative zone in their network. Every 6th bi-weekly release, representing a major release, is then deployed to the entire network as illustrated in Figure \ref{fig:DevOpsSWflow}.  

On the services side, the case study company offers several product-related services to its traditional customers, such as deployment, customer support, optimization, and education. Each service is considered \textit{stand-alone}, as customers can select and contract what suits them. The introduction of the bi-weekly releases has led the case study company to introduce a new set of services for the collaboration customers. While these services are performed as part of the collaboration agreement without a commercial framework, the case study company aims to commercialize these services.  

This study focuses on the role of services with collaborating customers for two reasons. First, since a frequent and reliable release process is one of the key enablers of DevOps, the collaborators are the ones deploying the two-week release cycle. Second, as the \textit{collaborator} name indicates, the case study company and these customers consider its relationship collaborative, which is a key characteristic of DevOps. 

%% file: content/5results.tex
\section{Empirical Findings} \label{devservops_results}

This section presents the findings from the case study company. Based on the analysis of the data collected during our two years participant observation case study, it can be summarized that services play a role on two fronts: the Dev front by providing \textit{internal} services, and the Ops by providing \textit{external} services. The role of the internal services is to: (1) help customers speed up the deployment of new software and (2) continuously extract value from the software to help customers meet their business outcomes. Internal services aim to (1) collect feedback data and (2) create actionable insights.  

\subsection{External services}

Once a release is delivered to customers, several operational activities shall be performed, which need to happen quickly as the time between each consecutive release is short. These activities include customer lab validation, new software impact analysis, pre-deployment preparation, deployment, post-deployment monitoring, and new feature validation. 

\begin{figure}[htbp]
	\begin{center}
	\includegraphics[width=3.0in]{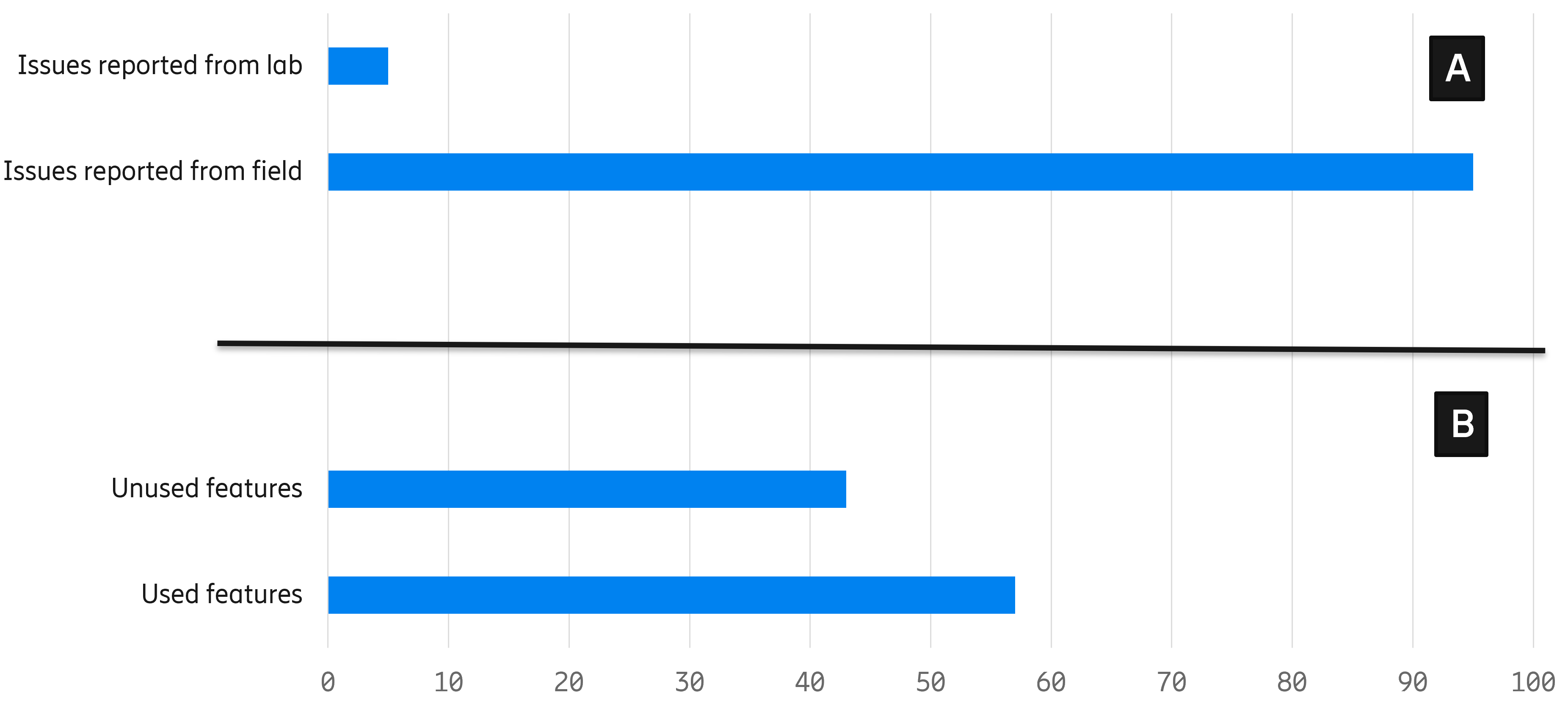}
	\caption{A. Issues found in the lab versus the field, B. Used versus unused software features}
	\label{fig:ratios_devservops}
	\end{center}
\end{figure}

Lab validation aims to verify that the new software works in a similar environment to the customer's live network. While lab validation is considered an important step for many operators before the deployment, the number of software issues discovered is small in relation to the issues reported after the software is deployed into field nodes, as shown in Figure \ref{fig:ratios_devservops} (A). This is because customers' lab often contains simpler configurations to nodes in the field and lacks real mobile traffic with the same intensity and variations as generated by thousands of mobile terminals. However, in several meetings, it was highlighted that lab validation is still important to ensure that basic functionality, such as emergency call handling, works satisfactorily in the customer's environment, as many countries have local regulations to ensure reliable and accurate handling of emergency communications, such as the EU directive 2018/1972 \cite{EU:2018/1972}. 

Furthermore, every release comes with several new features, improvements, and limitations. Thus, customers need to understand the release's content and assess the software's impact before the deployment. Therefore, customers need to read the release documentation and review the release content list. A feature toggle often controls new features; thus, they will not be enabled by default once the software is deployed. However, features that are already activated require more careful understanding and preparation before the deployment, as improvements to these features introduced in a new software version will be activated directly as the feature toggle is already enabled. For example, a feature improvement might impact the Performance Indicator (KPI) values or introduce a sequence change for certain operational activities. Therefore, such changes should be identified before the deployment to prepare for the impact of the new software. 

The pre-deployment preparation and deployment procedure follows a lab validation and impact analysis. In preparation for the deployment, a backup from the gNodeBs is taken and the new software is uploaded to the nodes. The actual Software deployment is often performed during a low-traffic period, such as between midnight and 04:00 local time, when the nodes are restarted for the new software to be activated. Following that, configuration checks are performed to ensure that non-default and customized configuration values used by customers are configured, followed by a health check to ensure that the gNodeBs are operating without any alarms or errors. Following the upgrade, the gNodeBs are continuously monitored for at least two to three days to ensure no traffic degradation is observed during high-traffic hours, such as early mornings and afternoons. After that, new features validation and trials start. Customers select features of interest and evaluate them. This activity also involves impact analysis and continuous monitoring. Further, when there is a risk that a new feature might introduce a negative impact, its activation and evaluation are performed during a low-traffic period. 

Therefore, external services are associated with customers' operational activities and have two main roles:  (1) enable faster deployment and (2) extract tangible business value to help customers reach their business outcomes.

\subsubsection{Enabling faster deployments}

While the operational activities need to happen with every software deployment, despite how frequent the release cycle is, the short-release cycle requires that these activities happen faster to meet the software's release frequency. As highlighted during several meetings, the upgrade procedure involves many steps which take time and effort. Therefore, there are three services that help customers with frequent deployments: testing, flow management, intelligent troubleshooting, and fault isolation. 

Testing services are aimed at helping customers cope with fast-paced releases and their lab tests. The service involves human resources and testing infrastructure, where service engineers deploy and integrate the testing automation infrastructure in the customer's lab. In addition, the service engineers work closely with the customer's operation team to perform the customer's testing scope while also advising them on how to optimize the testing scope.  
    
Further, flow management services aim at simplifying the planning and execution of the upgrades according to the customer's calendar. In several meetings, it was highlighted that customers have certain times when no changes are allowed to the network, known as a network freeze period, which happens during public holidays and other national occasions. Therefore, the services team, composed of technical and project management resources, works closely with the customer to plan the execution of the upgrades according to the customer's calendar. For example, the services team can perform lab validation when a network freeze is active and perform the impact analysis in preparation for faster software deployment once the freeze period is over. If the freeze period is longer than two weeks, which means that two more releases are waiting for deployment, the service team assesses the impact of each release and advice the customer if they should deploy both releases or drop one release due to time limitations. 
    
Intelligent troubleshooting and fault isolation aim at helping customers quickly and efficiently find issues that impact the network and minimize any potential impact. The services organization is engaged continuously with the customers' operations team in discussing identified deviations and finding their root causes. Intelligent troubleshooting and fault isolation rely on continuous product data collection from the customer network, where performance, diagnostic, and observability data are collected. The collected data is fed to a monitoring platform, which has the ability to identify deviations in the 5G RAN nodes' behavior and alert the service team to take further steps. In addition, the monitoring platform has embedded AI/ML models that can proactively fix known issues or predict their occurrence based on historical data. Utilizing the data flow, historical data, and AI/ML techniques, this service minimizes the risks associated with introducing new software while ensuring the continuity and reliability of the 5G RAN network. 

\subsubsection{Extracting the value of software}

One of the challenges discussed in several meetings is customers' difficulty in identifying relevant features and quantifying their value as the step enabling new features adoption. Providing a new release and documentation describing the release's features and content does not mean customers will use them. As illustrated in part B of Fig. \ref{fig:ratios_devservops}, many features are not used by any customer. Therefore, two services are identified to help customers identify relevant features and quantify their value: new features discovery and consultation, in addition to devitalized collaboration and benchmarking.

New features discovery and consultation services aim to simplify the impact analysis of the new software by working closely with the customer to review the release content and advise customers of relevant features that suit their business objectives. Customers often have different business objectives connected to how they position themselves in the market. For example, some customers are focused on performance, therefore, are interested in features that improve the network performance, while others might be interested in energy efficiency, therefore, interested in features optimizing energy consumption. Thus, the services team advises customers on features that suit their business objectives and work closely with them to validate them, quantify their impact, and ensure that customers enable them across the whole network. In addition, the service team provides consultation security consultations by working closely with customers to ensure security recommendations are implemented according to the guidelines associated with the release documentation. Similarly, the services team advises and guides the customer to implement any workaround due to known limitations in the release, such as known bugs that are not yet corrected.

Furthermore, services also enable collaboration between customers by providing a digitalized collaboration forum where customers' operation team members can engage, discuss and share best practices and observations with each other. In this forum, the service team acts as facilitators and administrators for the forum rather than the ones answering the questions. Customers can contact the services team for dedication questions or support requests via the service contact center. The forum has certain rules that shall be followed to ensure that the information shared is not sensitive, not harmful, and technically accurate. In addition to the forum, the services team provides an anonymized benchmarking service that helps customers to benchmark their network's metrics with other customers based on performance data collected continuously from each customer. Therefore, customers can compare their network to the best in class in different categories

\subsection{Internal services}

Internal services are aimed at the development organization and its functions, such as software development, product management, continuous integration, testing, and release engineering. Based on the case study company's findings, the internal services' role is to collect feedback and create actionable insights. 

\subsubsection{Data collection}

Services play a key role in collecting product data and customer feedback. To achieve this, the case study company establish a data collection pipeline that connects to the radio nodes in the customer's network and continuously collects performance and diagnostic data. Due to country-specific regulations and the characteristics of the data, and customer preference, the collected data can be transferred back to the case study company and shared across different locations within the case study company's internal network, or it can be hosted in the same geography in case there are regulations limiting cross-borders transfer of data. The service organization discusses and agrees upon the data collection pipeline technical solution with the customer. As a result, the service organization provides and maintains a number of customized data collection solutions to satisfy different regulations and customer preferences.  

In addition to product data, the service organization collects feedback from customers due to its close interaction with the customer's operations teams. The feedback is conveyed verbally, over emails, or via the digitalized collaboration platform. Customer feedback provides the service organization with a closer insight into how customers perceive aspects such as usability, simplicity, and clarity of documentation.  

\subsubsection{Actionable insights}

In several meetings and discussions, it was highlighted that the amount of information collected is huge, which makes it very difficult for the development organization to utilize it in an actionable way without putting so much effort into trying to understand the context of the feedback and data. For example, if product data shows an increase in the number of alarms during a certain period of time, would that indicate a software fault that requires a root cause identification and a correction, or is it because the customer's operation team were performing a configuration activity on the radio node. Similarly, if the customer report that certain functionality does not provide the expected benefit as described in the feature guide, would that be because the feature is not working as it should, or the customer has made a configuration mistake, or the pre-conditions that should exist for the feature to work are not present in the radio node. 

Therefore, the service organization uses both customer feedback and product-generated data to provide \textit{actionable insights} to the development organization. This is achieved by cleaning, aggregating, and qualification of the feedback and product data and then presenting them in an actionable way to different users within the development organization.

%% file: content/6discussion.tex
\section{DevServOps} \label{devservops_discussion}

DevOps advocates for a strong collaboration between development and operations to streamline the activities from a code change until that code is realized in the production environment. While DevOps is widely adopted in companies developing web-based applications, the context is different for companies producing software-intensive products. These companies do not only provide products to their customers but a bundle of products and product-related services aimed at fulfilling customers' needs. Product-related services are critical to ensure customer satisfaction besides their major contribution to companies' revenue. Thus, customer service is often the interface of the company towards its customers, delivering services such as deployment, support, and maintenance \cite{gebauer2016services}. In addition, in companies developing web-based applications, the Dev and Ops are often represented by different teams or organizations within the same company. However, in the case of software-intensive products, the \textit{Ops} is the customer who uses and operates the product. Further, software-intensive products are often high volume; thus, unlike web-based applications with few Ops teams and a limited number of production environments, there are many customers and product instances in the case of software-intensive products \cite{lwakatare2016towards}. 

\begin{figure}[htbp]
	\begin{center}
	\includegraphics[width=3.2in]{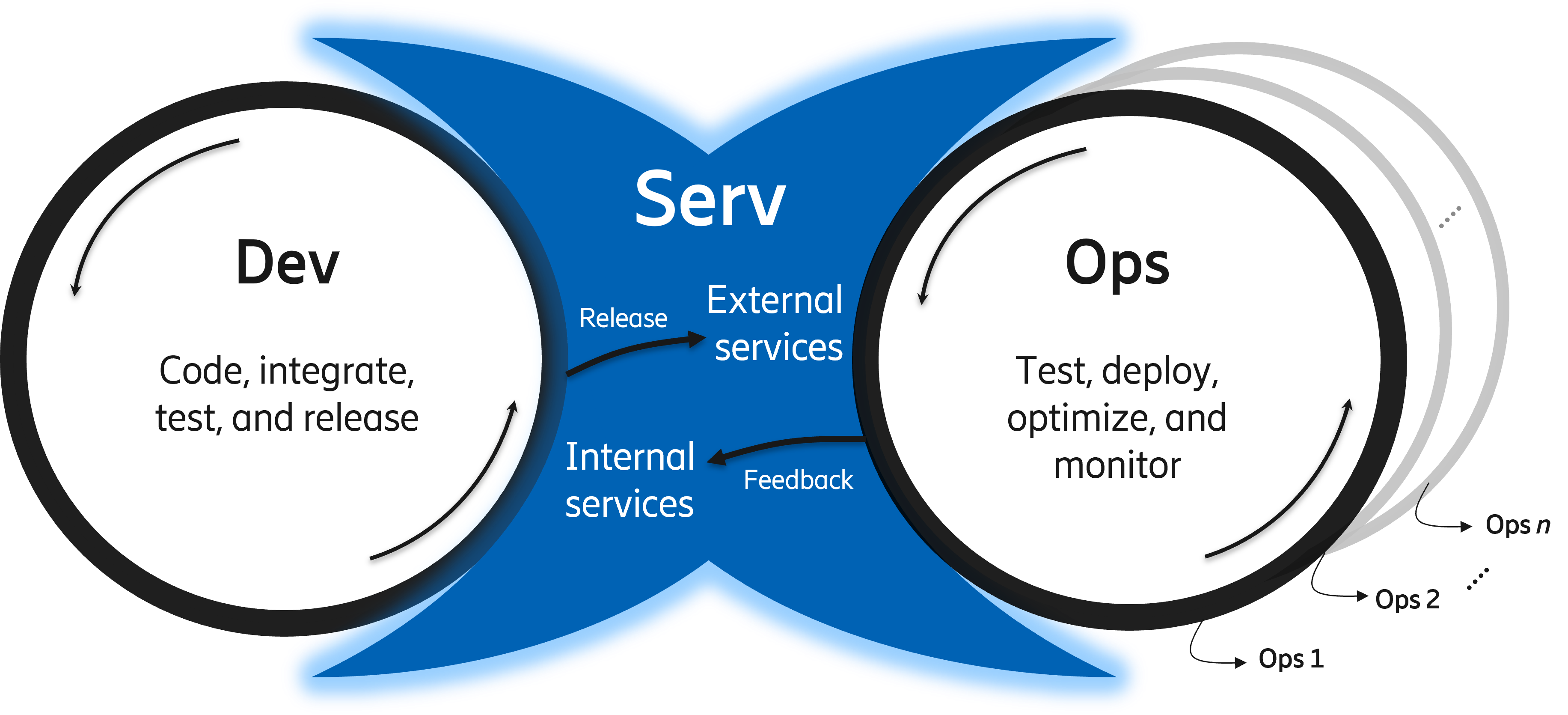}
	\caption{DevServOps}
	\label{fig:DevServOps}
	\end{center}
\end{figure}

Therefore, to address the context of product-oriented PSS companies, we propose a new and novel approach called Development-Services-Operations (DevServOps) to emphasize the role of services as the glue connecting the Dev and the Ops to streamline the end-to-end flow of the software to customers on one direction, and the feedback to developers on the other direction. As depicted in Figure \ref{fig:DevServOps}, the service organization is represented by \textbf{serv} and interfaces both the product development organization hosting the software development teams, represented by Dev, and the customers, represented by Ops. Several Ops instances are depicted to reflect the high volume nature of software-intensive products operated by different customers, ranging from 1 to \textit{n}. 

In DevServOps, services work on two fronts: the Dev and the Ops. On the Dev front providing internal services by collecting customer feedback from product data and customer interactions. As highlighted by Van et al. \cite{van2021role} customer services is a major channel for feedback to product development. In addition, to make such feedback actionable, the service organization cleans and aggregates the data, then extracts relative insights and presents that to different actors in the development organization. This enables the development organization to focus on what matters and protect development resources from being exposed to noisy information, which does not add value but rather wastes their time and effort.

On the customer front, services play a key role in helping customers cope with the fast pace releases by providing testing and flow management services to help them validate the new software in their environment and managing the increased frequency of deployments. In addition, services help to ensure that new features are discovered by customers and tangible customer value is extracted. As highlighted by Claps et al. \cite{claps2015journey}, the frequent delivery of new software increases the risk that new features are not discovered by customers. In addition, as the software complexity increases as a consequence of more features being introduced in the code leading to more complex interactions between the features, even if not used \cite{fabijan2016time}, the service organization plays a key role to minimizes the risks associated with introducing new software by working closely with customers utilizing intelligent and efficient troubleshooting and faults isolation capabilities. 

%% file: content/7conclusion.tex
\section{Conclusion} \label{devservops_conclusion}

DevOps is widely adopted in companies developing web-based applications as the collaboration between the Dev and Ops is key to streamlining the software flow. However, companies producing software-intensive products have a different context, as these companies do not only provide products to their customers but Product Service Systems, a bundle of products and services aimed at fulfilling customers' needs. Product-related services play a critical role in ensuring customer satisfaction as services are often the interface towards customers who operate and use the products, besides their major revenue contribution. 

Therefore, this paper investigated the role of services with DevOps in product-oriented PSS. Based on a two-year participant observation case study, we propose a new and novel approach called Development-Services-Operations (DevServOps) which incorporates services as a key activity to facilitate an end-to-end software flow toward customers in one direction and feedback flow toward developers in the other direction. Thus, DevServOps provides a holistic approach that addresses product-oriented PSS companies' context and ensures customer satisfaction.